\documentclass[letterpaper,journal]{IEEEtran}
\usepackage{amsmath,amsfonts}
\usepackage[caption=false,font=normalsize,labelfont=sf,textfont=sf]{subfig}
\usepackage{stfloats}
\usepackage{graphicx}
\usepackage{cite}
\usepackage{multirow}
\usepackage{booktabs}
\usepackage[switch]{lineno}
\usepackage[hidelinks]{hyperref}
\usepackage[OT1]{fontenc} 
\usepackage{tikz}
\newcommand\copyrighttext{%
  \footnotesize \textcopyright \the\year{} IEEE. Personal use of this material is permitted. Permission from IEEE must be obtained for all other uses, including reprinting/republishing this material for advertising or promotional purposes, collecting new collected works for resale or redistribution to servers or lists, or reuse of any copyrighted component of this work in other works.}

\newcommand\copyrightnotice{%
\begin{tikzpicture}[remember picture,overlay]
\node[anchor=south,yshift=10pt] at (current page.south) {\fbox{\parbox{\dimexpr0.75\textwidth-\fboxsep-\fboxrule\relax}{\copyrighttext}}};
\end{tikzpicture}%
}

\renewcommand\fbox{\fcolorbox{red}{white}}
\def\BibTeX{{\rm B\kern-.05em{\sc i\kern-.025em b}\kern-.08em
T\kern-.1667em\lower.7ex\hbox{E}\kern-.125emX}}
\begin{document}

\title{MuSTRNG: A Muon-Based Synchronizable True Random Number Generation System}

\author{Bohang Zou, Shubin Liu, Yu Wang, Ting Wang, Changqing Feng, and Zhongtao Shen%
\thanks{This work has been submitted to the IEEE for possible publication.
Copyright may be transferred without notice, after which this version may no longer be accessible.}%
\thanks{This work was supported by the National Science Fund for Distinguished Young Scholars (Grant No. 12025504)  
and the Young Scientists Fund of the National Natural Science Foundation (Grant No. 12205297).
(Corresponding author: Shubin Liu.)}%
\thanks{Bohang Zou, Yu Wang, Ting Wang, and Zhongtao Shen are with the Department of Modern Physics,
University of Science and Technology of China, Hefei 230026, China.}%
\thanks{Shubin Liu and Changqing Feng are with the Department of Modern Physics,
University of Science and Technology of China, Hefei 230026, China,
and also with the National Key Laboratory of Deep Space Exploration,
University of Science and Technology of China, Hefei 230026, China
(e-mail: liushb@ustc.edu.cn).}%
}

\maketitle
\copyrightnotice

\begin{abstract}
Cosmic-ray muons provide naturally occurring stochastic events detectable by spatially separated devices, offering a potential physical source for synchronized random-bit generation. This is possible because the same physical muons can be independently detected by both devices and used to derive corresponding random-bit segments. Therefore, identifying common muon events and correctly aligning the event sequences are essential for synchronized random-bit generation. This article presents MuSTRNG, a muon-based synchronizable true random number generation system with two synchronization methods: Tracking--Tracking and Tracking--Nontracking. For Tracking--Tracking synchronization, track-based screening removes events whose reconstructed tracks do not intersect the detector area of the other device; angle-based screening partitions reconstructed directions into angular bins and aligns the event sequences to remove inconsistent pairs; and timestamp-based screening compares higher-order timestamp bits to further suppress residual mismatches. Tracking--Nontracking synchronization applies track-based screening only at the tracking-capable device, followed by timestamp-based screening between the two devices. A prototype with two tracking-capable devices was evaluated. For Tracking--Tracking synchronization, the three-stage process reduced the invalid-event ratio to 0.28\% using 15-bit random segments, while the generated bit sequences showed generally favorable results in the applicable NIST SP~800-22 tests. Reprocessing the data without track information from one device provided preliminary support for Tracking--Nontracking synchronization, yielding an invalid-event ratio of 0.30\% using 5-bit random segments. These results demonstrate the feasibility of muon-based synchronized random-bit generation under different device configurations, with potential application to secure communications.
\end{abstract}


\section{Introduction}

\IEEEPARstart{C}{osmic-ray} muons are secondary charged particles produced primarily by the decay of mesons generated in extensive air showers initiated by interactions of high-energy primary cosmic rays with the atmosphere \cite{takahashi2026,gaisser2016}. Cosmic-ray muons constitute a major component of the secondary cosmic radiation at ground level and exhibit several distinctive properties, including high penetrating power, velocities close to the speed of light, and a persistent natural flux \cite{takahashi2026,gaisser2016}. Their arrival directions exhibit a characteristic angular distribution, while tracking detectors enable their trajectories to be reconstructed on an event-by-event basis. Owing to these properties, cosmic-ray muons are widely used in science and engineering, including detector calibration and characterization, muographic imaging, geophysical investigations, and inspection of dense or shielded materials \cite{bonomi2020,bonechi2020}.

Beyond these established applications, cosmic-ray muons possess additional characteristics that make them attractive as a physical source for generating shared random information. In particular, their arrival times exhibit stochastic behavior \cite{ahlen1992,becherini2005}, providing a naturally occurring source of temporal randomness. Moreover, owing to their strong penetrability, a single muon can traverse multiple spatially separated detectors positioned along its trajectory, allowing the same physical event to be detected at multiple locations. Cosmic-ray muons near sea level have a mean energy of approximately 4~GeV \cite{workman2022}, corresponding to a velocity of approximately 99.97\% of the speed of light. Thus, for time-of-flight estimation, the muon propagation velocity can be approximated as the speed of light, and the difference between the arrival times recorded at different detector locations is determined primarily by the muon time of flight. Consequently, when the same muon event is independently detected and timestamped by multiple devices, the corresponding measurements can be used to derive shared random information without requiring an artificial radiation source or dedicated wired connection.

\begin{figure}[!t]
\centering
\includegraphics[width=3.4in]{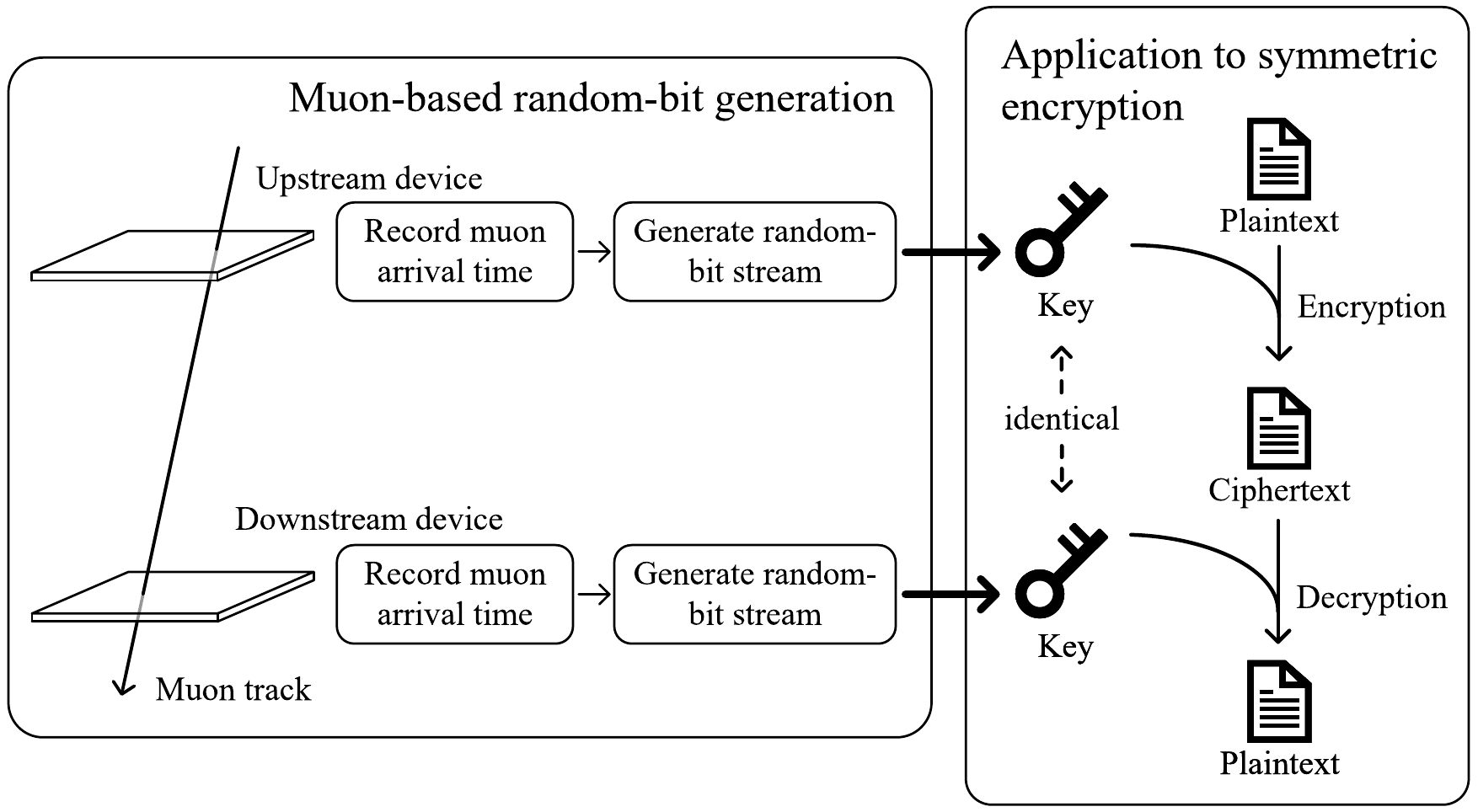}
\caption{Principle of muon-based random-bit generation and its potential application to symmetric encryption.}
\label{Fig1}
\end{figure}

These characteristics have motivated previous studies on the use of cosmic-ray muon arrival times for random-number generation and cryptographic applications \cite{tanaka2023muon_nfc,tanaka2023muon_key_storage,tanaka2025muon_password_gen}. As illustrated in Fig.~\ref{Fig1}, when the same muon is detected by spatially separated devices, the recorded arrival times can be converted into random bit streams from which identical keying material may be derived for symmetric encryption. Symmetric-key cryptography offers high computational efficiency and relatively low communication overhead, making it well suited for secure data transmission \cite{menezes2018}. In symmetric encryption, the sender and receiver share the same secret key: the sender uses it to encrypt plaintext, while the receiver uses it to recover the plaintext. Accordingly, secure key establishment and assurance that both parties possess the same keying material are important for symmetric-key cryptographic systems \cite{barker2020}. Muon-based approaches therefore offer the potential to derive such shared keying material from naturally occurring physical events without requiring a dedicated wired connection between the devices. Moreover, the continual arrival of cosmic-ray muons offers the possibility of periodically refreshing the derived keys.

However, a fundamental limitation of previously reported muon-based key-generation systems is the lack of an explicit event-screening mechanism for reliably determining whether independently recorded events at different devices originate from the same physical muon \cite{tanaka2023muon_nfc,tanaka2023muon_key_storage,tanaka2025muon_password_gen}. For spatially separated detectors, only a subset of the muon events recorded by either device are common to both devices, while events recorded by only one device can lead to mismatched random outputs. Existing systems generally accommodate such mismatches through repeated trials with candidate keys or passwords during subsequent encryption or authentication \cite{tanaka2023muon_nfc,tanaka2023muon_key_storage,tanaka2025muon_password_gen}. This approach becomes less effective as the detector separation increases or as more muon events are combined to generate a key, because the probability that all constituent events are correctly matched decreases, resulting in a lower probability of successful key or password matching and an increased number of required trials. This limitation motivates the development of effective methods for identifying common muon events and aligning the corresponding event sequences to generate synchronized random outputs.

In this work, we propose MuSTRNG, a Muon-Based Synchronizable True Random Number Generation system. MuSTRNG exploits reconstructed tracks, incident angles, and partial timestamp information to identify commonly detected muons and align the corresponding event sequences. Synchronization methods are developed for both Tracking--Tracking and Tracking--Nontracking device configurations. A prototype system is constructed to acquire experimental muon data, which are subsequently processed offline to evaluate Tracking--Tracking synchronization and the statistical randomness of the generated bit sequences. The experimental data are further reprocessed with the track information from one device excluded to provide a preliminary assessment of Tracking--Nontracking synchronization.

The remainder of this article is organized as follows. Section II introduces the principle of muon-based random number generation. Section III describes the proposed synchronization methods. Section IV describes the experimental prototype and presents the experimental results. Section V discusses practical deployment and security considerations. Finally, Section VI concludes this article.

\section{Principle of Muon-Based Synchronizable Random Number Generation}

\subsection{Generation of Identical Random Bits}
For brevity, cosmic-ray muons are hereafter referred to as ``muons.'' Each device employs a detector and a free-running counter driven by a measurement clock with period $T_{\mathrm{clk}}$. When a muon is detected, the current counter value $n$ is latched as its timestamp:
\begin{align}
n = \left\lfloor\frac{t-t_0}{T_{\mathrm{clk}}}\right\rfloor,
\end{align}
where $t$ is the muon arrival time and $t_0$ denotes a common reference epoch for the counters, which is assumed to be established beforehand using an auxiliary synchronization mechanism.

\begin{figure}[!t]
\centering
\includegraphics[width=3.2in]{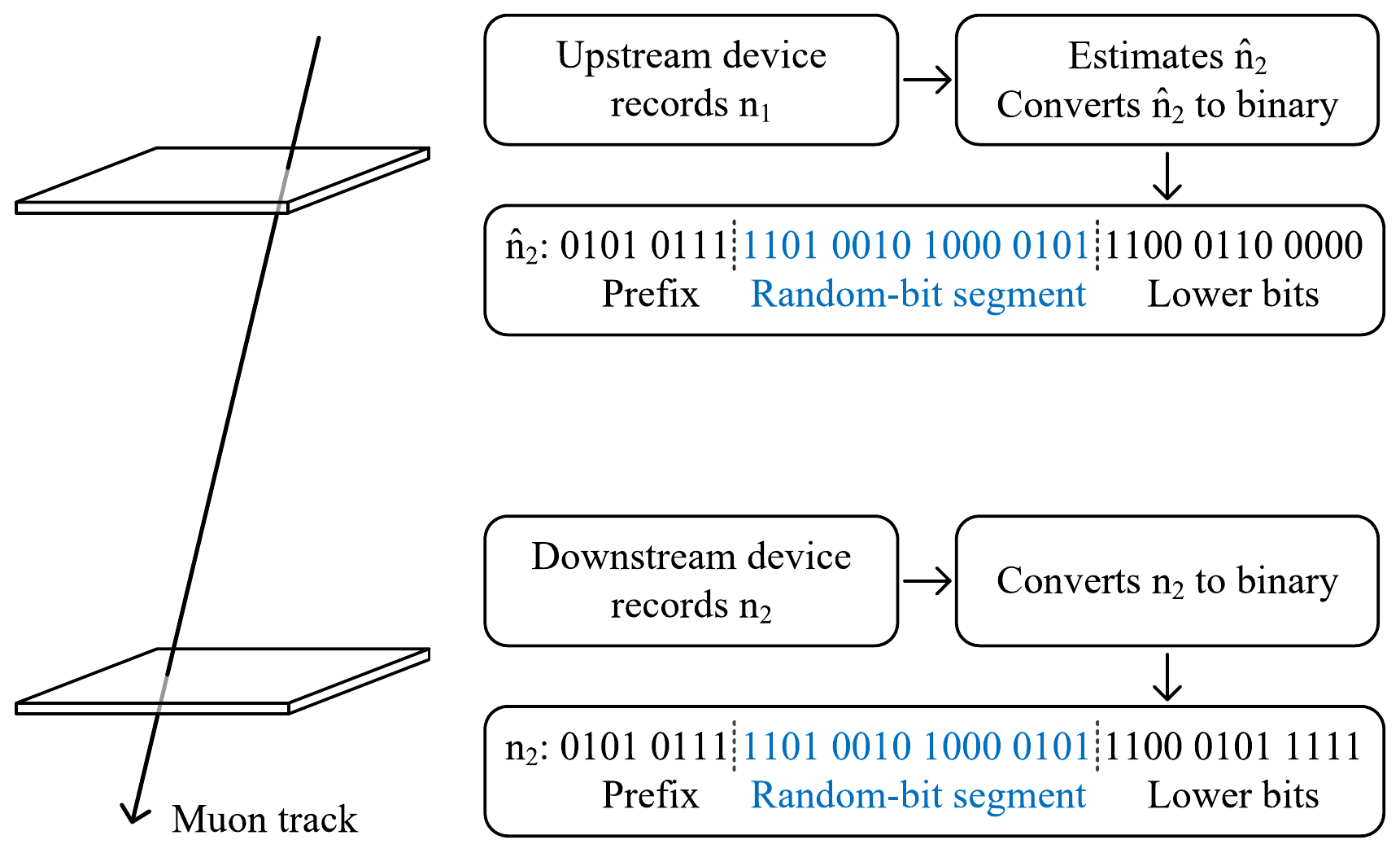}
\caption{Generation of identical random bits from a muon detected by both devices.}
\label{Fig2}
\end{figure}

For a muon detected by both devices, the upstream and downstream devices record timestamps $n_1$ and $n_2$, respectively. The principle of generating identical random bits from these timestamps is illustrated in Fig.~\ref{Fig2}. Approximating the propagation velocity of the relativistic muon as $c$, the upstream device estimates the downstream timestamp $\hat{n}_2$ as
\begin{align}
\hat{n}_2 = n_1 + \operatorname{round}\left(\frac{\ell_\mu}{cT_{\mathrm{clk}}}\right) = n_2 + \delta n,
\end{align}
where $\ell_\mu$ is the muon propagation path length between the two detectors and $\delta n = \hat{n}_2-n_2$ denotes the residual timestamp difference.

The two devices then convert the corresponding timestamps, $\hat{n}_2$ and $n_2$, into binary form and extract a $k$-bit segment from predetermined bit positions shown in Fig.~\ref{Fig2}. The higher-order bits preceding the selected segment, collectively referred to as the prefix, change relatively slowly over consecutive events and therefore provide limited event-to-event variation. In contrast, the lower-order bits are more sensitive to the residual timestamp difference $\delta n$ and are consequently more difficult to synchronize. The segment position and length $k$ are selected to preserve the randomness arising from stochastic muon arrival times while reducing the probability that $\delta n$ causes different $k$-bit segments at the two devices. With an appropriate selection, the two devices can generate identical $k$-bit segments from the same muon event with high probability. For the device configurations considered here, $k$ typically ranges from several bits to tens of bits. The same procedure can be applied in the reverse direction, with the downstream device estimating $\hat{n}_1$ from $n_2$ and the two devices extracting the same predetermined bit positions from $\hat{n}_1$ and $n_1$, respectively.

\subsection{Construction of Synchronized Random Bit Strings}

A single $k$-bit segment is generally too short for practical cryptographic use; for example, the minimum key length specified for the Advanced Encryption Standard (AES) is 128 bits \cite{fips197}. Multiple $k$-bit segments must therefore be concatenated to construct a random bit string of the required length. The muon events can be selected according to a predefined rule, such as using consecutive events in chronological order.

For a pair of devices, muon events detected by both devices are hereafter referred to as ``valid events,'' whereas events detected by only one device are referred to as ``invalid events.'' If all events used for bit-string construction are valid and correctly associated between the two devices, the corresponding $k$-bit segments can be concatenated in the same order to produce identical random bit strings. These identical bit strings can subsequently be used as shared input keying material for the two devices, from which identical symmetric keys can be derived using a key-derivation function, such as the HMAC-based Extract-and-Expand Key Derivation Function (HKDF) \cite{rfc5869}.

With a fixed number of bits extracted from each valid event, the achievable random-bit rate is determined by the valid-event rate. Under otherwise comparable conditions, this rate generally increases with larger detector areas, higher detection efficiencies, and smaller detector separations, while also depending on the relative detector geometry and orientation. However, synchronized random-bit generation also requires reliable discrimination between valid and invalid events.

\subsection{Limitation of Existing Muon-Based Schemes}

The preceding principle assumes that the muon events used for bit-string construction are valid and correctly associated between the two devices. In practice, however, the event sequences independently recorded by spatially separated devices generally contain invalid events because only a fraction of the muons detected by one device are also detected by the other. Such invalid events may enter the sequences used for bit-string construction, causing bit segments derived from different physical muon events to be combined and thereby producing mismatched bit strings and, subsequently, inconsistent symmetric keys.

This problem becomes more pronounced as the device separation increases or as longer random outputs are required. Increasing the separation generally reduces the fraction of muons detected by both devices, whereas constructing a longer output requires a larger number of correctly associated valid events. Therefore, reliable identification and alignment of valid muon events are essential for improving the synchronization of muon-derived random bit strings. This limitation motivates the screening and alignment methods introduced in MuSTRNG.

\section{Synchronization Methods}

MuSTRNG combines the timestamp-based random-bit generation described in Section II with event screening and sequence alignment to construct synchronized random-bit strings between spatially separated devices. To achieve reliable event correspondence without exchanging the extracted random-bit segments themselves, MuSTRNG exploits auxiliary information associated with the detected events, including reconstructed tracks, incident angles, and partial timestamp information.

\begin{figure}[!t]
\centering
\includegraphics[width=2.8in]{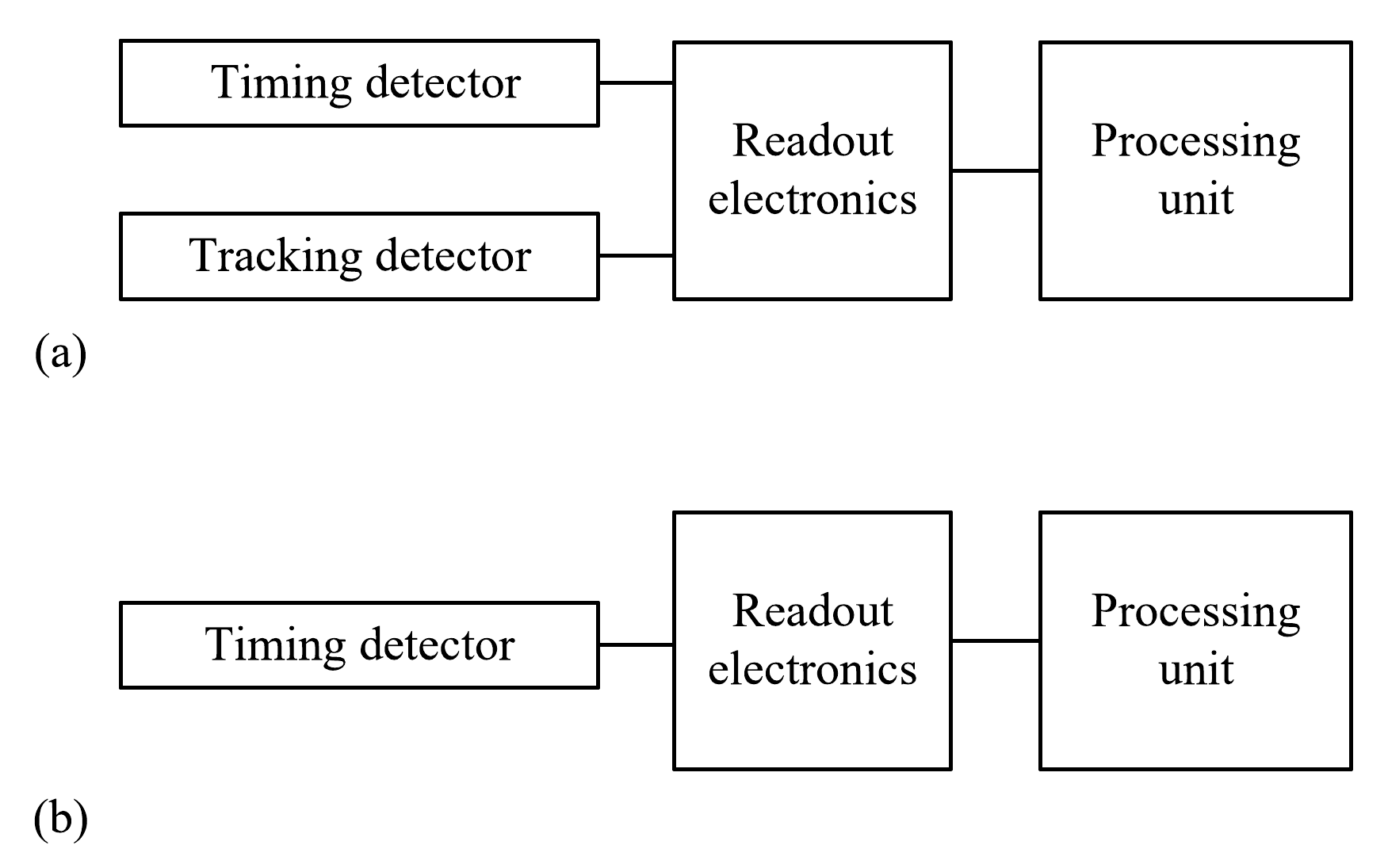}
\caption{Device configurations considered in MuSTRNG: (a) tracking-capable device and (b) nontracking device.}
\label{Fig3}
\end{figure}

According to the detector capabilities available at the two devices, two synchronization methods are considered. As illustrated in Fig.~\ref{Fig3}, a tracking-capable device comprises a timing detector, a tracking detector, readout electronics, and a processing unit. The timing detector provides the timing signal for timestamp generation, whereas the tracking detector provides the information required for track reconstruction. In contrast, a nontracking device does not include a tracking detector and therefore provides only timing information for each detected muon event. The readout electronics acquire the detector signals and generate event timestamps, while the processing unit performs subsequent event processing and synchronization.

Accordingly, MuSTRNG supports Tracking--Tracking synchronization, in which track information is available at both devices, and Tracking--Nontracking synchronization, in which track information is available at only one device. In either configuration, the relative detector geometry should provide sufficient acceptance for muons to be detected by both devices.

\subsection{Tracking--Tracking Synchronization}

Tracking--Tracking synchronization consists of three stages: track-based screening, angle-based screening, and timestamp-based screening. Both devices provide track and timing information to perform all three screening stages. The three stages progressively reject invalid events and refine the correspondence between the event sequences recorded by the two devices. The retained events are subsequently used to generate synchronized random-bit outputs.

\begin{figure}[!t]
\centering
\includegraphics[width=2.7in]{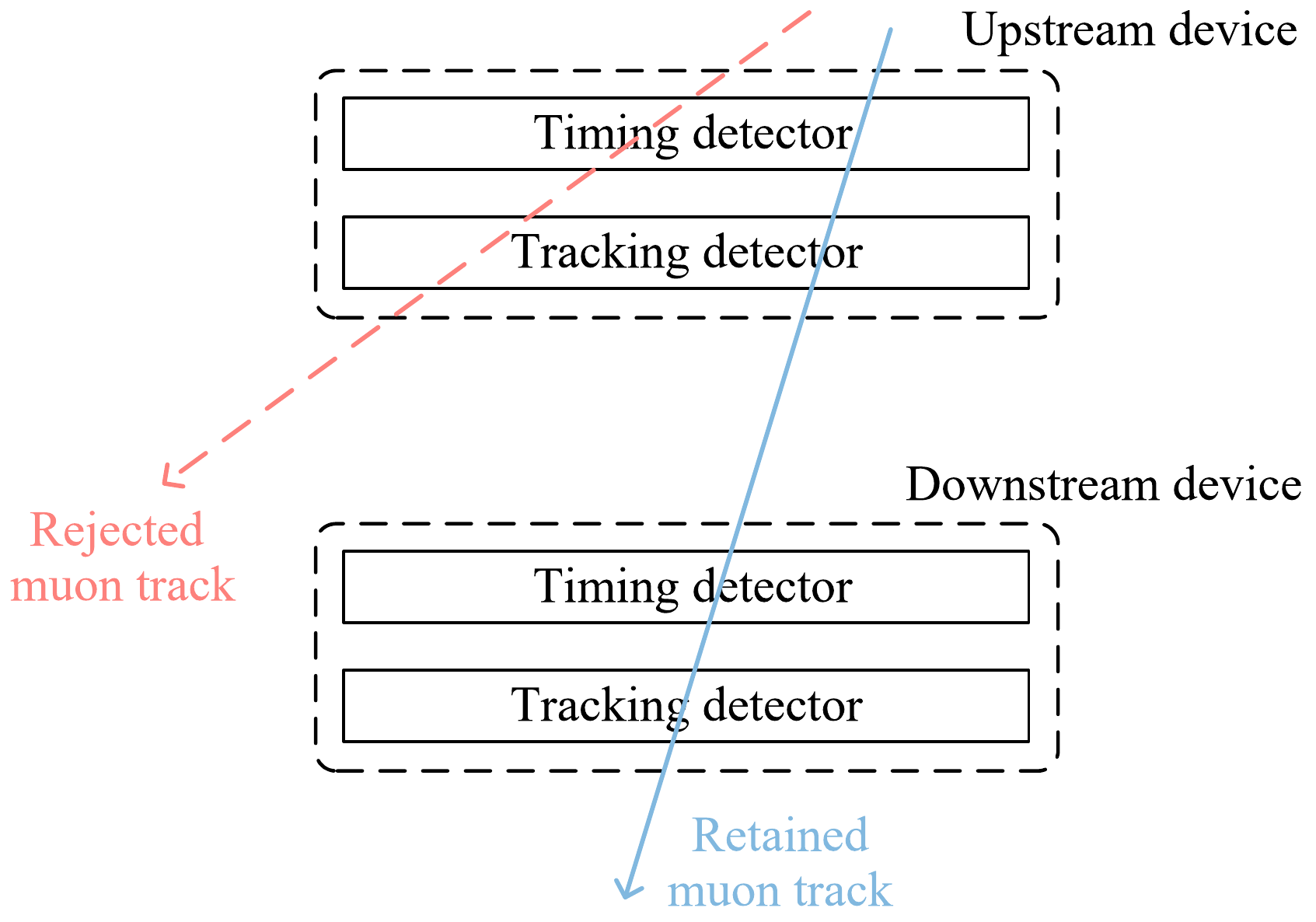}
\caption{Principle of track-based screening in Tracking--Tracking synchronization.}
\label{Fig4}
\end{figure}

The first stage is track-based screening. As illustrated in Fig.~\ref{Fig4}, each device reconstructs the tracks of its detected muons from the tracking-detector information. Based on the known relative positions and geometries of the two devices, an event is retained only if its reconstructed track, when extrapolated between the devices, intersects the required detector areas at both devices; otherwise, it is discarded. Both devices independently perform this screening before the subsequent stages.

The second stage is angle-based screening, which further removes invalid events resulting primarily from missed detections at either device due to finite detection efficiency. The reconstructed tracks from the two devices are transformed into a common coordinate system using their known relative positions and orientations. The direction of each reconstructed track is characterized by its zenith angle $\theta$ and azimuth angle $\phi$. For the same muon event, the directions reconstructed independently by the two devices should differ only within a small angular range. This range is mainly determined by the angular resolution of the tracking detectors and by muon scattering along the propagation path. This limited angular deviation provides a basis for identifying corresponding events in the two event sequences.

After track-based screening, the retained tracks lie within the common angular acceptance of the two devices. The allowable ranges of $\theta$ and $\phi$ are determined from the known detector geometry and uniformly partitioned into $N_{\theta}$ and $N_{\phi}$ intervals, respectively. Both devices use the same angular ranges and bin boundaries. Each retained event is then assigned a discrete angular label $(a,b)$, where $a$ and $b$ denote the bin indices of $\theta$ and $\phi$, respectively. As events are accumulated, each device constructs a time-ordered sequence of angular labels. Each label remains associated with the $k$-bit segment generated from the corresponding event, so that retaining, removing, or aligning an event results in the corresponding operation on its associated $k$-bit segment.

When the same muon event is detected by both devices, the corresponding angular labels are expected to be identical or sufficiently close. In contrast, if an event is missed by one device because of finite detection efficiency, it appears only in the sequence recorded by the other device, introducing a gap between the two sequences. Each sequence element is represented by a two-dimensional angular label $(a,b)$. Two labels are regarded as a match when the differences between their zenith- and azimuth-angle bin indices are within predefined tolerances, with the periodicity of the azimuth angle taken into account.

\begin{figure}[!t]
\centering

\subfloat[]{%
    \includegraphics[width=0.9\columnwidth]{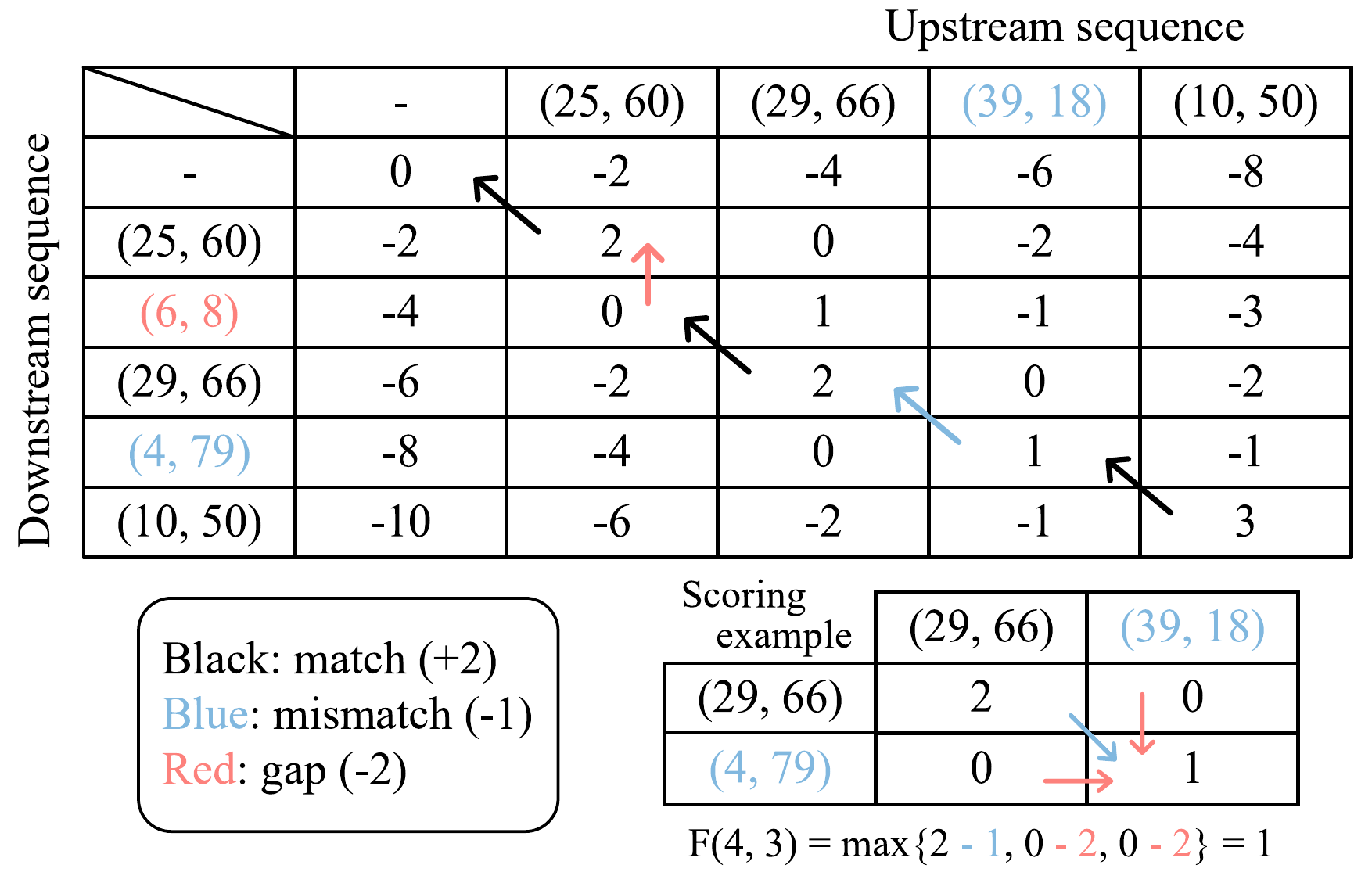}%
    \label{fig:angle_score_matrix}}

\par\vspace{0.5mm}

\subfloat[]{%
    \includegraphics[width=0.95\columnwidth]{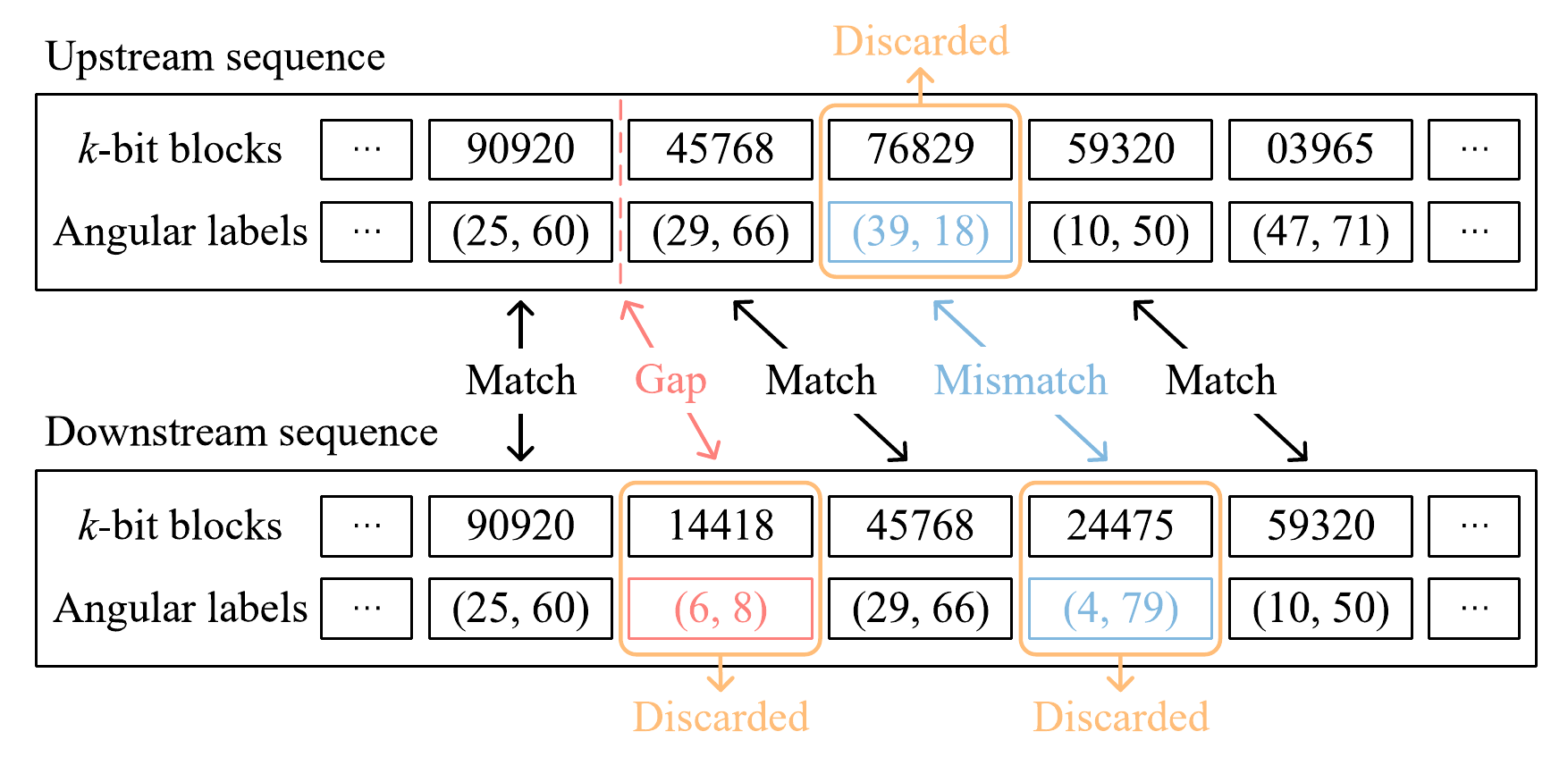}%
    \label{fig:angle_alignment}}
 
\caption{Example of angular-label sequence alignment.
(a) Scoring matrix and traceback of the modified Needleman--Wunsch (NW) algorithm. The inset illustrates the calculation of $F(4,3)$, where the diagonal mismatch and the two gap candidates yield scores of $1$, $-2$, and $-2$, respectively, giving $F(4,3)=1$. (b) Resulting sequence alignment, where events associated with gaps and mismatches, together with their corresponding $k$-bit segments, are discarded.}
\label{Fig5}
\end{figure}

To identify gaps and mismatches between the two sequences, the devices exchange their angular-label sequences and perform global sequence alignment using a modified Needleman--Wunsch (NW) algorithm \cite{needleman1970}. For two sequences of lengths $M$ and $N$, a scoring matrix $F$ of size \mbox{$(M+1)\times(N+1)$} is constructed. With a gap score of $-2$, the matrix is initialized as $F(0,0)=0$, $F(i,0)=-2i$, and $F(0,j)=-2j$. Each remaining element is calculated as
\begin{align}
F(i,j) = \max\left\{
\begin{array}{l}
F(i-1,j-1) + s(i,j),\\
F(i-1,j) - 2,\\
F(i,j-1) - 2,
\end{array}
\right.,
\end{align}
where $s(i,j)=2$ if the two angular labels satisfy the matching criterion and $s(i,j)=-1$ otherwise. Thus, the match, mismatch, and gap scores are $2$, $-1$, and $-2$, respectively. During matrix construction, each element is assigned a traceback direction corresponding to the predecessor that yields the maximum score. When multiple candidate moves yield the same score, the priority order is match, upward gap, leftward gap, and mismatch. After the matrix is completed, traceback starts from the bottom-right element $F(M,N)$ and follows the assigned directions to obtain the global sequence alignment.

An example of the scoring matrix and traceback procedure is shown in
Fig.~\ref{Fig5}\subref{fig:angle_score_matrix}, and the resulting sequence alignment
is illustrated in Fig.~\ref{Fig5}\subref{fig:angle_alignment}. A diagonal traceback step aligns one event from each sequence and is classified as either a match or a mismatch according to the angular-label matching criterion, whereas an upward or leftward step introduces a gap corresponding to an event present in only one sequence. Events associated with gaps or mismatches, together with their corresponding $k$-bit segments, are discarded, while matched events are retained for the subsequent screening stage.

The third stage is timestamp-based screening, which further removes mismatched events that survive angle-based screening because their angular labels coincidentally satisfy the matching criterion. As defined in Section II, the timestamp prefix consists of the higher-order bits preceding the selected $k$-bit segment. For a valid event, the two devices are expected to obtain the same timestamp prefix. Since timestamp prefixes form nondecreasing sequences during an acquisition run, their temporal ordering and repetition patterns can be exploited to identify and remove residual mismatches.

\begin{figure}[!t]
\centering
\includegraphics[width=3in]{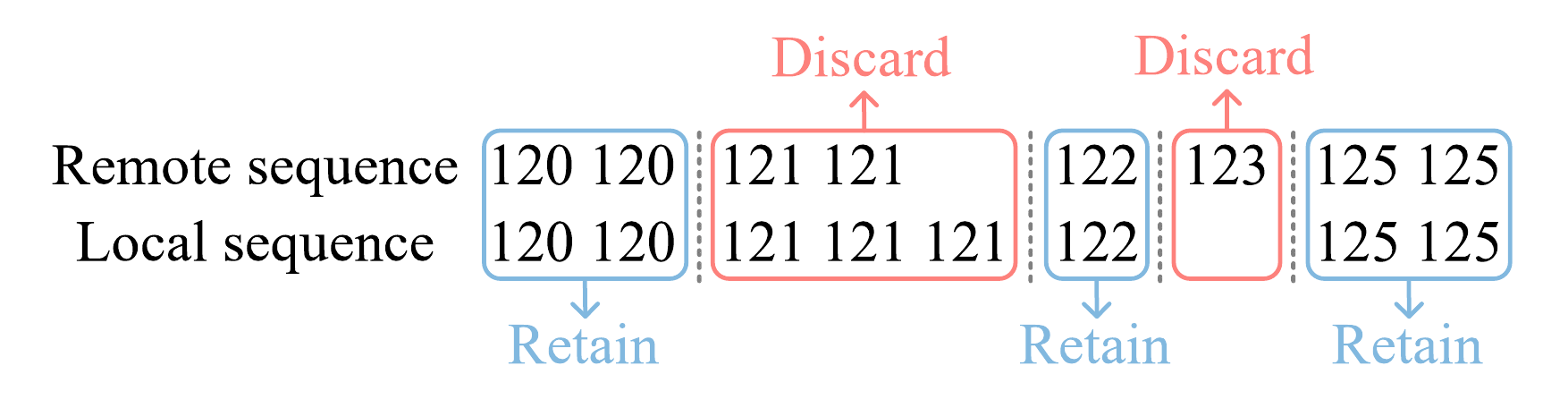}
\caption{Timestamp-prefix sequence screening based on prefix ordering and group size.}
\label{Fig6}
\end{figure}

Each device stores the timestamp prefixes associated with the events retained after angle-based screening and exchanges the resulting prefix sequence with the other device. Consecutive events sharing the same prefix are treated as a prefix group, and the two nondecreasing sequences are compared group by group. As illustrated in Fig.~\ref{Fig6}, a prefix group is retained when the same prefix appears the same number of times in both sequences. If a prefix occurs in only one sequence, the corresponding events are discarded. If the same prefix appears in both sequences but a different number of times, the event correspondence within that group is ambiguous, and the entire group is conservatively discarded. By exploiting both prefix ordering and group size, this procedure further suppresses residual mismatches and improves the correspondence between the retained event sequences. 

Because the exchanged angular labels are derived from reconstructed track information and the exchanged timestamp prefixes do not overlap with the selected $k$-bit segments, the exchanged auxiliary information does not include the generated $k$-bit segments themselves.

\subsection{Tracking--Nontracking Synchronization}

To simplify the secondary device and reduce its hardware requirements, Tracking--Nontracking synchronization removes the tracking detector from one of the two devices, allowing it to operate as a nontracking endpoint. Tracking--Nontracking synchronization therefore consists of two stages: track-based screening and timestamp-based screening. Since track information is available only at the tracking-capable device, the track-based stage is performed exclusively by this device, whereas the nontracking device relies solely on its recorded arrival-time information. The events retained after the two stages are subsequently used to generate synchronized random-bit outputs.

In the first stage, the tracking-capable device reconstructs the tracks of its detected muons and evaluates them based on the known relative geometry of the two devices. The screening criterion is analogous to that used in Tracking--Tracking synchronization, except that it is applied only at the tracking-capable device. An event is retained if its reconstructed track, when extrapolated to the nontracking device, intersects the sensitive detector area; otherwise, it is discarded.

The second stage follows the same timestamp-based screening procedure described for Tracking--Tracking synchronization. After the track-based stage, the two devices exchange the timestamp-prefix sequences associated with the events entering this stage. The two nondecreasing sequences are compared according to prefix ordering and group size: groups with the same prefix value and the same number of occurrences in both sequences are retained, whereas unmatched groups or groups with ambiguous event correspondence are conservatively discarded. Unlike Tracking--Tracking synchronization, only the sequence at the tracking-capable device has undergone track-based screening before the prefix comparison. The resulting sequences have identical timestamp-prefix patterns and lengths, and residual invalid events are further suppressed. 

Because only timestamp prefixes are exchanged between the two devices and these prefixes do not overlap with the selected $k$-bit segments, the exchanged auxiliary information does not include the generated $k$-bit segments themselves.

\section{Prototype and Experimental Results}

\subsection{Prototype}

\begin{figure}[!t]
\centering
\includegraphics[width=3.4in]{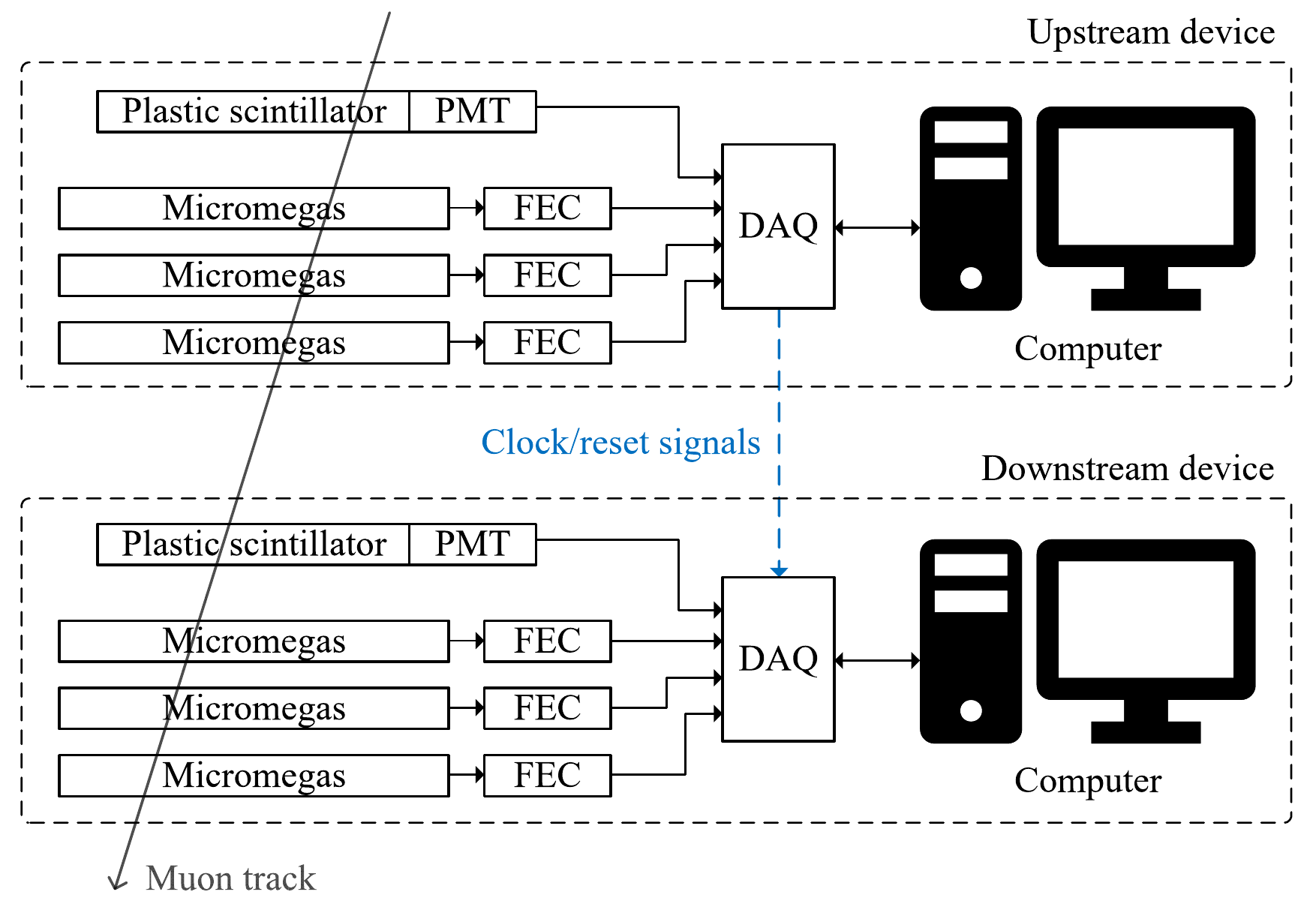}
\caption{Architecture of the experimental prototype.}
\label{Fig7}
\end{figure}

An experimental prototype was constructed to acquire muon data for evaluating the proposed MuSTRNG synchronization methods. As shown in Fig.~\ref{Fig7}, the prototype consists of two identical tracking-capable devices, referred to as the upstream and downstream devices. Each device includes a plastic scintillator detector, three Micromegas detectors, three front-end electronics cards (FECs), a fiber-based data acquisition (DAQ) board, and a computer.

Muon arrival times were measured using a 40~cm $\times$ 40~cm square plastic scintillator coupled to a photomultiplier tube (PMT; Hamamatsu R7525). Signals produced by muon traversals were read out by the PMT and transmitted to the DAQ board through a coaxial cable.

A 200-MHz clock on the DAQ board drives a free-running counter implemented in a Xilinx Kintex-7 FPGA (XC7K410T-FFG900-2L), corresponding to a clock period of $T_{\mathrm{clk}}=5$~ns. The PMT signal is amplified and fed to a comparator. When the amplified signal exceeds a preset threshold, the current counter value is latched as the event timestamp.

The Micromegas detectors, each with a 60~cm $\times$ 60~cm active area, were used for muon-track measurement. A Micromegas detector records the two-dimensional hit position of a charged particle. In each device, three such detectors were arranged sequentially, providing multiple hit positions from which the muon track was reconstructed.

The Micromegas detectors used in this study were fabricated using a thermal-bonding method, and their fabrication process and performance have been reported in \cite{feng2021}. Although Micromegas detectors provide high spatial resolution, their large number of readout channels increases system cost, power consumption, and data-processing complexity. To reduce readout complexity, an encoding circuit developed in our previous work \cite{wang2025}, which employs a position-encoded multiplexing method, was connected between each Micromegas detector and its corresponding FEC.

The FECs process the Micromegas signals and communicate with the DAQ board through optical fiber links. The DAQ board distributes synchronized clock and trigger signals to the FECs, receives data streams from them, and performs data framing and packetization for transmission to the PC. The FECs \cite{zhu2019} and the DAQ board \cite{liu2025daq}, developed in our previous work, constitute a versatile readout system for high-granularity muography detectors. Detailed descriptions of the readout architecture and performance are provided in \cite{liu2025sys}.

Coincidence-trigger logic was implemented in the FPGA on the DAQ board to suppress unrelated background noise. Within each device, an event was recorded, packetized, and transmitted to the computer only when the plastic scintillator and all three Micromegas detectors generated trigger signals within a predefined coincidence time window of $5~\mu\mathrm{s}$. The received packets were then decoded on the computer to extract the event timestamp and hit-position information for muon-track reconstruction. The resulting timestamps and reconstructed tracks were stored chronologically for subsequent offline analysis.

The two devices were synchronized using a shared clock transmitted over differential signal lines and a common reset signal transmitted through a coaxial cable. The shared clock eliminated accumulated relative drift between the two counters, while the common reset established a nominal common reference epoch for both counters. Any remaining fixed timestamp offset between the devices was calibrated during offline data processing.

\begin{figure}[!t]
\centering
\includegraphics[width=2.4in]{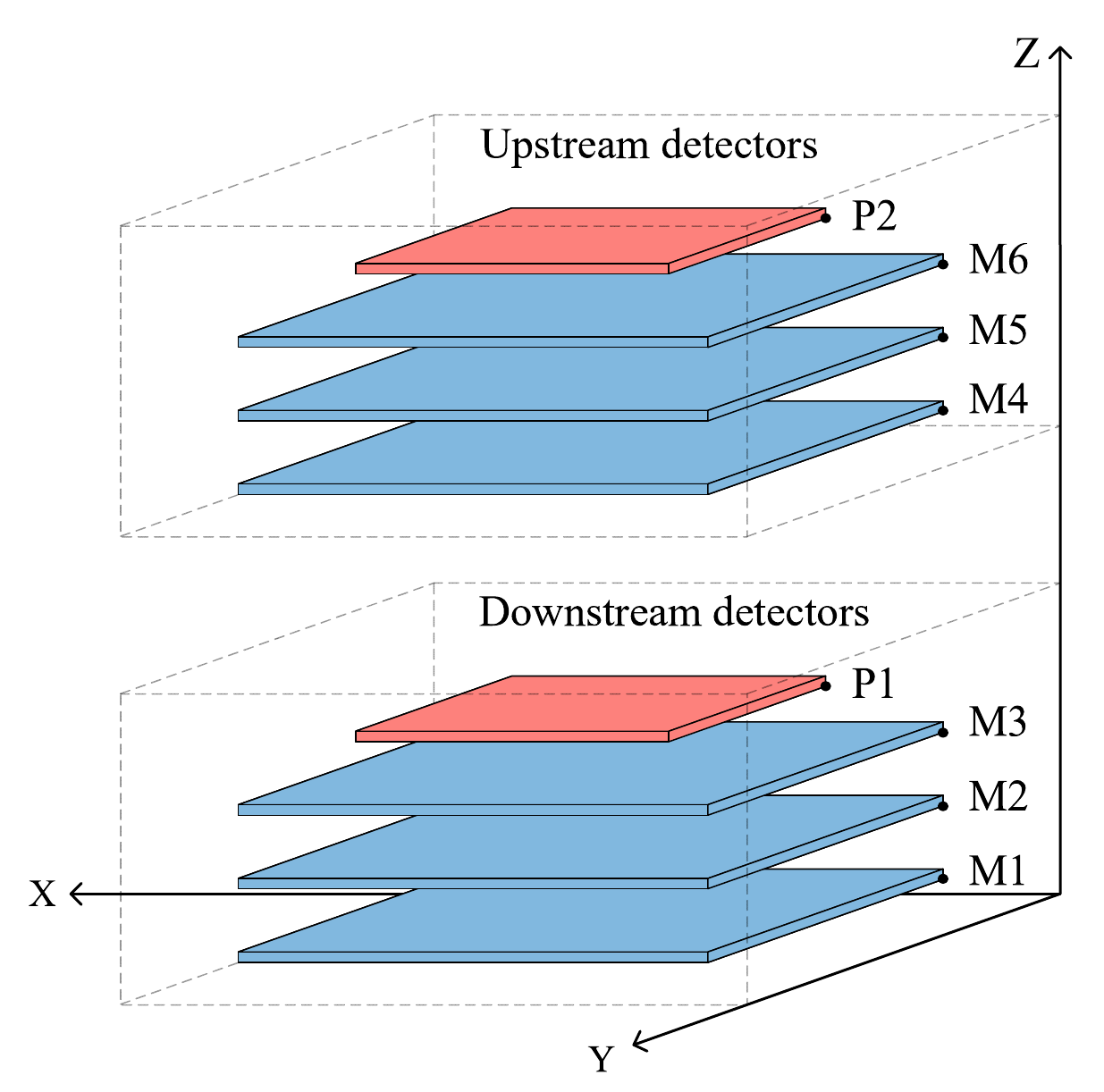}
\caption{Detector arrangement of the experimental prototype. The red plates represent plastic scintillators, and the blue plates represent Micromegas detectors.}
\label{Fig8}
\end{figure}

The relative arrangement of the detectors in the experimental prototype is illustrated in Fig.~\ref{Fig8}. The detector planes were arranged approximately parallel to one another, with negligible relative rotation, and their centers were nearly aligned along the vertical axis. The vertical positions of the detectors relative to M1 are summarized in Table~\ref{tab:detector_height}.

\begin{table}[!t] 
\caption{Detector Heights Relative to M1} 
\label{tab:detector_height} 
\centering 
\begin{tabular}{c c} 
\hline Detector & Height Relative to M1 (mm) \\ 
\hline P2 & 1302 \\
M6 & 1218 \\
M5 & 1068 \\
M4 & 909 \\
P1 & 382 \\
M3 & 305 \\
M2 & 163 \\
M1 & 0 \\
\hline 
\end{tabular} 
\end{table}

\subsection{Experimental Results}

The prototype was operated in four independent runs, each lasting 20~h. Since both devices were equipped with tracking detectors, the acquired data were first used to evaluate the proposed three-stage screening process for Tracking--Tracking synchronization. The statistical randomness of the resulting synchronized bit sequences was also examined. In addition, the same experimental data were reprocessed to provide a preliminary evaluation of Tracking--Nontracking synchronization by omitting the tracking information from the second device.

\subsubsection{Evaluation of Tracking--Tracking Synchronization}

After correcting the fixed timestamp offset between the two devices, each timestamp was represented in binary form. Bit indices are counted from the least significant bit (LSB), starting from 1. In the experiments, bits 1--11 were discarded, bits 12 through $11+k$ were selected as the $k$-bit random segment, and bits above $11+k$ were used as the timestamp prefix for timestamp-based screening. With a 200-MHz timestamp clock, the least significant bit of the selected segment corresponds to a time interval of $2^{11}\times5$~ns, or approximately $10.24~\mu\mathrm{s}$.

To evaluate event correspondence throughout the screening process, the corrected timestamps were compared within each 10-min acquisition interval after removing the 11 least significant bits. Before sequence alignment, an event was regarded as matched if the same remaining timestamp value appeared in the sequence recorded by the other device. After angle-based screening established one-to-one correspondence between the two equal-length sequences, events at the same sequence index were compared pairwise, and a pair was regarded as matched only if the two remaining timestamp values were identical.

Track-based screening was performed first, and the event statistics before and after this stage are summarized in Table~\ref{tab:track_screening_statistics}. For either device, the invalid-event ratio among the retained events was calculated as
\begin{equation}
r_{\mathrm{invalid},i}
= 1 - \frac{n_{\mathrm{mat}}}{n_i},
\qquad i \in\{\mathrm{up},\mathrm{down}\},
\end{equation}
where $n_i$ denotes the number of retained events at the corresponding device and $n_{\mathrm{mat}}$ denotes the number of matched events identified by the a posteriori timestamp comparison. Here, $r_{\mathrm{invalid}}$ is an operational estimate based on timestamp consistency rather than an independent event-by-event verification that two recorded events originate from the same physical muon.

\begin{table}[!t]
\caption{Event Statistics for Track-Based Screening}
\label{tab:track_screening_statistics}
\centering
\footnotesize
\setlength{\tabcolsep}{3.2pt}
\renewcommand{\arraystretch}{1.08}
\setlength{\aboverulesep}{0pt}
\setlength{\belowrulesep}{0.3pt}
\begin{tabular}{c c c c c c c}
\hline
\multirow{2}{*}{\shortstack{Test\\No.}} &
\multirow{2}{*}{Stage} &
\multicolumn{3}{c}{Event count} &
\multicolumn{2}{c}{$r_{\mathrm{invalid}}$ (\%)} \\
\cmidrule(lr){3-5}\cmidrule(lr){6-7}
& & Up & Down & Matched & Up & Down \\
\hline
\multirow{2}{*}{1}
  & Original & 215,619 & 254,425 & 13,134 & --    & --    \\
  & Track    &  33,015 &  35,985 & 10,297 & 68.81 & 71.39 \\
\hline
\multirow{2}{*}{2}
  & Original & 235,254 & 261,244 & 14,504 & --    & --    \\
  & Track    &  36,064 &  36,879 & 11,365 & 68.49 & 69.18 \\
\hline
\multirow{2}{*}{3}
  & Original & 239,140 & 263,833 & 15,152 & --    & --    \\
  & Track    &  36,093 &  37,324 & 11,770 & 67.39 & 68.47 \\
\hline
\multirow{2}{*}{4}
  & Original & 217,382 & 263,178 & 14,102 & --    & --    \\
  & Track    &  32,875 &  37,527 & 10,960 & 66.66 & 70.79 \\
\hline
\multirow{2}{*}{Total}
  & Original & 907,395 & 1,042,680 & 56,892 & --    & --    \\
  & Track    & 138,047 &   147,715 & 44,392 & 67.84 & 69.95 \\
\hline
\end{tabular}
\end{table}

Across the four tests, track-based screening reduced the retained event counts from 907,395 to 138,047 at the upstream device and from 1,042,680 to 147,715 at the downstream device, while retaining 44,392 of the 56,892 matched events identified in the original data. The resulting invalid-event ratios were 67.84\% and 69.95\% for the upstream and downstream devices, respectively. These results demonstrate that track-based screening substantially reduces the number of candidate events, although a considerable fraction of invalid events remains, motivating the subsequent angle-based screening.

After track-based screening, angle-based screening was performed using the reconstructed muon directions. Based on the common angular acceptance of the two devices, the zenith-angle range $\theta\in[0,0.5]~\mathrm{rad}$ and the azimuth-angle range $\phi\in[-\pi,\pi]$ were uniformly partitioned into $N_{\theta}$ and $N_{\phi}$ bins, respectively. For angular-label matching, the bin-index tolerances in the modified NW algorithm were set to 2 for both $\theta$ and $\phi$, such that two indices were regarded as matching when their absolute difference was less than 2, i.e., when they were identical or differed by one bin. The first and last azimuth bins were additionally treated as adjacent to account for the periodic boundary. Finer angular binning generally improves the discrimination between events with different reconstructed directions. However, because the reconstructed angles of the same muon event can differ slightly between the two devices, finer binning converts a given angular difference into larger bin-index differences, which are more likely to exceed the predefined matching tolerances and cause a matched event to be discarded.

\begin{figure*}[!t]
\centering
\subfloat[]{\includegraphics[width=3.2in]{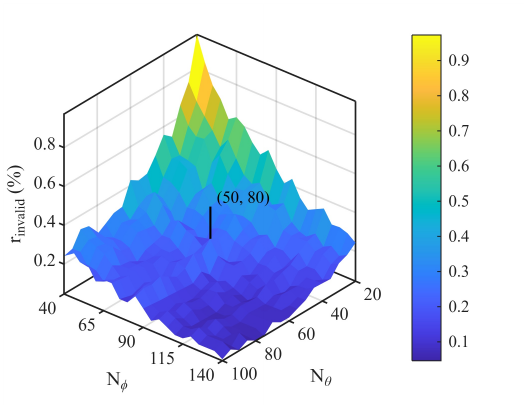}%
\label{fig:screen3_invalid_rate}}
\hfil
\subfloat[]{\includegraphics[width=3.2in]{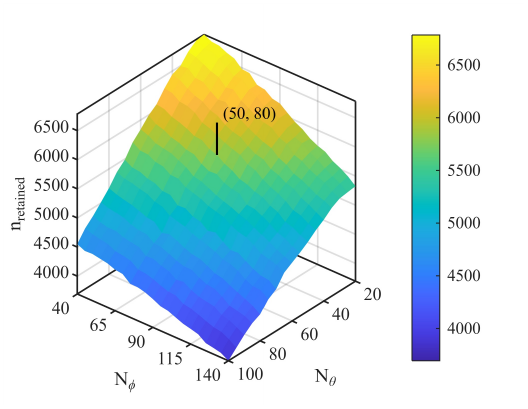}%
\label{fig:screen3_retained_events}}
\caption{Effects of $N_{\theta}$ and $N_{\phi}$ on the final screening results for Test No.~1 with $k=15$: (a) invalid-event ratio and (b) retained-event count.}
\label{fig:screen3_partition}
\end{figure*}

Because timestamp-based screening subsequently suppresses residual mismatches, excessively fine angular partitioning is unnecessary. To select suitable values of $N_{\theta}$ and $N_{\phi}$, both parameters were varied for Test No.~1 with $k=15$, and the final invalid-event ratio and number of retained events after all three screening stages were evaluated. As shown in Fig.~\ref{fig:screen3_partition}, increasing $N_{\theta}$ and $N_{\phi}$ generally lowers the invalid-event ratio but also reduces the number of retained events. Based on this tradeoff, $N_{\theta}=50$ and $N_{\phi}=80$ were selected and used for all four tests. If the detector area or relative geometry changes, the angular partition parameters should be re-optimized accordingly before synchronization.

After angle-based screening, the upstream and downstream sequences had equal lengths, and events at the same sequence index were compared pairwise. Timestamp-based screening preserved this one-to-one correspondence. Consequently, the two devices had identical retained-event counts and invalid-event ratios at both stages, and a single set of event statistics is reported for each stage in Table~\ref{tab:angle_timestamp_screening_statistics}.

\begin{table}[!t]
\caption{Event Statistics for Angle- and Timestamp-Based Screening With $k=15$}
\label{tab:angle_timestamp_screening_statistics}
\centering
\footnotesize
\setlength{\tabcolsep}{4.0pt}
\renewcommand{\arraystretch}{1.08}
\setlength{\aboverulesep}{0pt}
\setlength{\belowrulesep}{0.3pt}
\begin{tabular}{c c c c c}
\hline
\multirow{2}{*}{\shortstack{Test\\No.}} &
\multirow{2}{*}{Stage} &
\multicolumn{2}{c}{Event count} &
\multirow{2}{*}{$r_{\mathrm{invalid}}$ (\%)} \\
\cmidrule{3-4}
& & Retained & Matched & \\
\hline
\multirow{2}{*}{1}
  & Angle     & 6,199 & 5,685 & 8.29 \\
  & Timestamp & 5,687 & 5,675 & 0.21 \\
\hline
\multirow{2}{*}{2}
  & Angle     & 6,754 & 6,246 & 7.52 \\
  & Timestamp & 6,248 & 6,232 & 0.26 \\
\hline
\multirow{2}{*}{3}
  & Angle     & 7,079 & 6,559 & 7.35 \\
  & Timestamp & 6,566 & 6,548 & 0.27 \\
\hline
\multirow{2}{*}{4}
  & Angle     & 6,940 & 6,403 & 7.74 \\
  & Timestamp & 6,416 & 6,393 & 0.36 \\
\hline
\multirow{2}{*}{Total}
  & Angle     & 26,972 & 24,893 & 7.71 \\
  & Timestamp & 24,917 & 24,848 & 0.28 \\
\hline
\end{tabular}
\end{table}

The parameter $k$ also affects timestamp-based screening because it determines the number of higher-order timestamp bits available for prefix comparison. A smaller $k$ leaves a longer timestamp prefix, providing more timestamp information for distinguishing candidate events while yielding fewer random bits per retained event. To evaluate this tradeoff, the data from all four tests were reprocessed for $k=10$--$15$ with all other screening parameters fixed. For this evaluation, the synchronized-bit rate was defined as
\begin{equation}
R_{\mathrm{sync}}
=
\frac{k n_{\mathrm{mat}}}{T_{\mathrm{run}}},
\end{equation}
where $n_{\mathrm{mat}}$ is the number of matched event pairs after timestamp-based screening and $T_{\mathrm{run}}=80$~h is the total experimental run time of the four tests.

\begin{figure}[!t]
\centering
\includegraphics[width=3.2in]{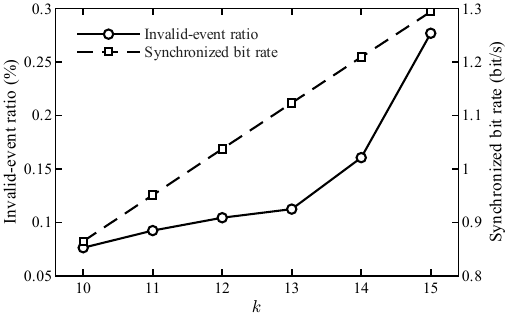}
\caption{Effects of $k$ on the invalid-event ratio and synchronized-bit rate for the combined data from the four tests.}
\label{Fig10}
\end{figure}

As shown in Fig.~\ref{Fig10}, the synchronized-bit rate increases with $k$. However, the invalid-event ratio also increases because a larger $k$ leaves fewer higher-order timestamp bits for prefix-based screening. This behavior reflects a tradeoff between synchronized-bit rate and screening performance. The value of $k$ can therefore be selected according to the requirements of a specific application.

Overall, the proposed three-stage screening process substantially improved the correspondence between the retained event sequences. With $N_{\theta}=50$, $N_{\phi}=80$, and $k=15$, the invalid-event ratios after timestamp-based screening ranged from 0.21\% to 0.36\% across the four tests. For the combined data, 24,848 of the 24,917 retained event pairs were identified as matched by the a posteriori timestamp comparison, corresponding to an invalid-event ratio of 0.28\%. These results demonstrate effective synchronization of the retained event sequences.

To evaluate the statistical randomness of the generated bit sequences, the NIST SP 800-22 Rev.~1a test suite \cite{rukhin2010} was applied with a significance level of $\alpha = 0.01$. Under the selected configuration with $N_{\theta}=50$, $N_{\phi}=80$, and $k=15$, the bit sequences generated by the upstream device from the four tests were concatenated. The first 24,915 complete 15-bit segments were partitioned into 55 sequences of 6,795 bits, with each sequence comprising 453 complete segments generated from retained muon events. The use of 55 sequences also satisfies the NIST recommendation for statistically meaningful evaluation of the uniformity of $P$-values \cite{rukhin2010}.

As summarized in Table~\ref{tab:nist_results}, the Frequency, Block Frequency, Runs, Longest Run, FFT, Approximate Entropy, Serial, and Cumulative Sums tests satisfied the corresponding passing criteria, while 144 of the 148 evaluated Non-overlapping Template subtests satisfied the passing-proportion criterion. Overall, the generated bit sequences showed generally favorable results in the applicable NIST SP 800-22 tests. Tests requiring substantially longer input sequences were not performed.

\begin{table}[!t]
\caption{NIST SP 800-22 Test Results for the Generated Bit Sequences}
\label{tab:nist_results}
\centering
\footnotesize
\setlength{\tabcolsep}{3.5pt}
\renewcommand{\arraystretch}{1.08}
\begin{tabular}{l c c c}
\hline
\multirow{2}{*}{Test Module} &
Uniformity &
\multirow{2}{*}{Pass Prop.} &
\multirow{2}{*}{Result} \\
& $P$-Value & & \\
\hline

Frequency
& 0.014550 & 55/55 & Pass \\

Block Frequency
& 0.275709 & 54/55 & Pass \\

Runs
& 0.554420 & 55/55 & Pass \\

Longest Run
& 0.514124 & 55/55 & Pass \\

FFT
& 0.249284 & 54/55 & Pass \\

Non-overlapping Template
& -- & -- & 144/148 passed$^{a}$ \\

Serial ($P_1$)
& 0.129620 & 54/55 & Pass \\

Serial ($P_2$)
& 0.759756 & 53/55 & Pass \\

Approximate Entropy
& 0.437274 & 55/55 & Pass \\

Cumulative Sums (Forward)
& 0.181557 & 55/55 & Pass \\

Cumulative Sums (Reverse)
& 0.304126 & 55/55 & Pass \\
\hline
\multicolumn{4}{l}{
\scriptsize $^{a}$144 of the 148 template subtests satisfied the passing-proportion criterion.
}
\end{tabular}
\end{table}

\subsubsection{Evaluation of Tracking--Nontracking Synchronization}

To provide a preliminary evaluation of Tracking--Nontracking synchronization, the data acquired in the four tests were reprocessed with the upstream device treated as the tracking-capable device and the downstream device treated as a nontracking device. Accordingly, the track information from the downstream device was excluded from the synchronization procedure. Track-based screening was performed only at the upstream device, after which the resulting upstream event sequence and the downstream event sequence were processed using timestamp-based screening.

As summarized in Table~\ref{tab:tracking_nontracking_track_statistics}, track-based screening reduced the upstream event count from 907,395 to 148,954 across the four tests, while retaining 47,773 of the 56,892 matched events. The downstream event sequence remained unchanged at this stage.

\begin{table}[!t]
\caption{Event Statistics for Track-Based Screening in
Tracking--Nontracking Synchronization}
\label{tab:tracking_nontracking_track_statistics}
\centering
\footnotesize
\setlength{\tabcolsep}{3.2pt}
\renewcommand{\arraystretch}{1.08}
\setlength{\aboverulesep}{0pt}
\setlength{\belowrulesep}{0.3pt}
\begin{tabular}{c c c c c c c}
\hline
\multirow{2}{*}{\shortstack{Test\\No.}} &
\multirow{2}{*}{Stage} &
\multicolumn{3}{c}{Event count} &
\multicolumn{2}{c}{$r_{\mathrm{invalid}}$ (\%)} \\
\cmidrule(lr){3-5}\cmidrule(lr){6-7}
& & Up & Down & Matched & Up & Down \\
\hline
\multirow{2}{*}{1}
  & Original & 215,619 & 254,425 & 13,134 & --    & --    \\
  & Track    &  35,611 & 254,425 & 11,127 & 68.75 & 95.63 \\
\hline
\multirow{2}{*}{2}
  & Original & 235,254 & 261,244 & 14,504 & --    & --    \\
  & Track    &  38,861 & 261,244 & 12,228 & 68.53 & 95.32 \\
\hline
\multirow{2}{*}{3}
  & Original & 239,140 & 263,833 & 15,152 & --    & --    \\
  & Track    &  38,976 & 263,833 & 12,695 & 67.43 & 95.19 \\
\hline
\multirow{2}{*}{4}
  & Original & 217,382 & 263,178 & 14,102 & --    & --    \\
  & Track    &  35,506 & 263,178 & 11,723 & 66.98 & 95.55 \\
\hline
\multirow{2}{*}{Total}
  & Original & 907,395 & 1,042,680 & 56,892 & --    & --    \\
  & Track    & 148,954 & 1,042,680 & 47,773 & 67.93 & 95.42 \\
\hline
\end{tabular}
\end{table}

\begin{figure}[!t]
\centering
\includegraphics[width=3.2in]{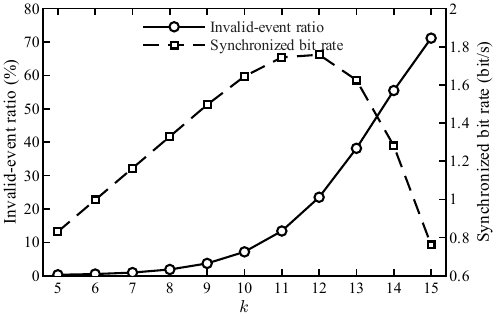}
\caption{Effects of $k$ on the invalid-event ratio and synchronized-bit rate for Tracking--Nontracking synchronization based on the combined data from the four tests.}
\label{Fig11}
\end{figure}

Timestamp-based screening was then performed. The effect of $k$ was further evaluated by reprocessing the combined data from the four tests with $k$ varied from 5 to 15. As shown in Fig.~\ref{Fig11}, the invalid-event ratio increases substantially with $k$, whereas the synchronized-bit rate first increases and then decreases. This behavior reflects a tradeoff between screening performance and synchronized-bit rate. At $k=5$, which favors synchronization quality, 47,773 of the 47,917 retained event pairs were identified as matched, corresponding to an invalid-event ratio of 0.30\% and a synchronized-bit rate of 0.83~bit/s. These results provide preliminary experimental support for the feasibility of the proposed Tracking--Nontracking synchronization method.

\section{Discussion}
The two synchronization configurations offer different tradeoffs between synchronization performance and detector complexity. In the present experiments, Tracking--Tracking synchronization achieved an invalid-event ratio of 0.28\% at $k=15$, while Tracking--Nontracking synchronization yielded an invalid-event ratio of 0.30\% at $k=5$. A larger $k$ requires fewer muon events to construct a bit string of a given length, thereby reducing the number of event associations required and the opportunity for association errors. In contrast, Tracking--Nontracking synchronization eliminates the tracking detector at one device, reducing hardware complexity, size, cost, and power consumption.

One possible application is an indoor security system. Tracking-capable devices can be deployed on different floors of a building as ``main nodes,'' whereas compact nontracking devices can be integrated into room-level endpoints such as smart locks, routers, cameras, and security gates as ``endpoint nodes.'' Main nodes can establish synchronized keying material using Tracking--Tracking synchronization, while each endpoint node can establish synchronized keying material with a nearby main node using Tracking--Nontracking synchronization. The resulting keying material can support symmetric-key secure communication, while communication between two endpoint nodes can be relayed through one or more main nodes.

Another practical consideration for distributed deployment is time synchronization. With independent clock sources, small frequency differences can cause the relative timestamp offset between the two devices to accumulate over time, potentially reducing the probability of generating identical $k$-bit segments from the same muon event. Wireless time-synchronization approaches, such as the cosmic time calibrator, have been investigated in related muon-based systems \cite{tanaka2025muon_password_gen,tanaka2023ctc}. In the present work, a shared clock was used to eliminate accumulated clock drift and allow the evaluation to focus on the proposed event-screening and sequence-alignment methods; this wired arrangement was adopted for controlled experimental evaluation rather than as a requirement of MuSTRNG. For practical deployments without a dedicated wired connection, a suitable time-synchronization mechanism would therefore be required.

From a security perspective, the two synchronization configurations may also present different security considerations. In Tracking--Tracking synchronization, reproducing the same keying material would generally require access to muon events correlated with those detected by the legitimate devices, together with the corresponding event associations. In Tracking--Nontracking synchronization, peer impersonation may warrant greater attention because the nontracking device lacks independent track information and therefore has fewer physical observables for checking the consistency of information received from its counterpart. Accordingly, peer authentication and protection of the exchanged auxiliary information are important topics for future work.

\section{Conclusion}

This article presented MuSTRNG, a muon-based synchronizable true random number generation system, together with the Tracking--Tracking and Tracking--Nontracking synchronization methods. A prototype with two tracking-capable devices was evaluated in four independent runs. For Tracking--Tracking synchronization, the proposed three-stage screening process reduced the invalid-event ratio to 0.28\% for the combined data at $k=15$, while the generated bit sequences showed generally favorable results in the applicable NIST SP~800-22 tests. Reprocessing the same data without using the downstream track information provided preliminary support for Tracking--Nontracking synchronization, yielding an invalid-event ratio of 0.30\% at $k=5$. These results demonstrate the feasibility of synchronized random-bit generation using cosmic-ray muons under both tracking and nontracking device configurations. Future work will focus on validation with a dedicated nontracking device, wireless time synchronization, and authentication and protection of exchanged auxiliary information.

\bibliographystyle{IEEEtran}
\bibliography{refs}

@article{takahashi2026,
  author  = {F. Takahashi and others},
  title   = {Review of particle physics},
  journal = {Int. J. Mod. Phys. A},
  volume  = {41},
  number  = {22},
  year    = {2026},
  note    = {Art. no. 2630011},
  doi     = {10.1142/S0217751X26300115}
}

@book{gaisser2016,
  author    = {T. K. Gaisser and R. Engel and E. Resconi},
  title     = {Cosmic Rays and Particle Physics},
  edition   = {2nd},
  address   = {Cambridge, U.K.},
  publisher = {Cambridge Univ. Press},
  year      = {2016}
}

@article{workman2022,
  author  = {R. L. Workman and others},
  title   = {Review of particle physics},
  journal = {Prog. Theor. Exp. Phys.},
  volume  = {2022},
  number  = {8},
  month   = aug,
  year    = {2022},
  note    = {Art. no. 083C01},
  doi     = {10.1093/ptep/ptac097}
}

@article{bonomi2020,
  author  = {G. Bonomi and P. Checchia and M. D'Errico and D. Pagano and G. Saracino},
  title   = {Applications of cosmic-ray muons},
  journal = {Prog. Part. Nucl. Phys.},
  volume  = {112},
  month   = may,
  year    = {2020},
  note    = {Art. no. 103768},
  doi     = {10.1016/j.ppnp.2020.103768}
}

@article{bonechi2020,
  author  = {L. Bonechi and R. D'Alessandro and A. Giammanco},
  title   = {Atmospheric muons as an imaging tool},
  journal = {Rev. Phys.},
  volume  = {5},
  month   = nov,
  year    = {2020},
  note    = {Art. no. 100038},
  doi     = {10.1016/j.revip.2020.100038}
}

@article{becherini2005,
  author  = {Y. Becherini and others},
  title   = {Time correlations of high energy muons in an underground detector},
  journal = {Astropart. Phys.},
  volume  = {23},
  number  = {3},
  pages   = {341--348},
  year    = {2005},
  doi     = {10.1016/j.astropartphys.2005.01.007}
}

@article{ahlen1992,
  author  = {S. P. Ahlen and others},
  title   = {Arrival time distributions of very high energy cosmic ray muons in {MACRO}},
  journal = {Nucl. Phys. B},
  volume  = {370},
  number  = {2},
  pages   = {432--444},
  month   = feb,
  year    = {1992},
  doi     = {10.1016/0550-3213(92)90292-J}
}

@techreport{barker2020,
  author      = {E. Barker},
  title       = {Recommendation for key management: Part 1---General},
  institution = {Nat. Inst. Standards Technol.},
  address     = {Gaithersburg, MD, USA},
  type        = {NIST Special Publication},
  number      = {800-57 Part 1 Rev. 5},
  month       = may,
  year        = {2020},
  doi         = {10.6028/NIST.SP.800-57pt1r5}
}

@article{tanaka2023muon_nfc,
  author  = {H. K. M. Tanaka},
  title   = {Cosmic coding and transfer for ultra high security near-field communications},
  journal = {iScience},
  volume  = {26},
  number  = {2},
  month   = feb,
  year    = {2023},
  note    = {Art. no. 105897},
  doi     = {10.1016/j.isci.2022.105897}
}

@article{tanaka2023muon_key_storage,
  author  = {{H. K. M. Tanaka}},
  title   = {Cosmic coding and transfer storage ({COSMOCATS}) for invincible key storage},
  journal = {Sci. Rep.},
  volume  = {13},
  month   = may,
  year    = {2023},
  note    = {Art. no. 8746},
  doi     = {10.1038/s41598-023-35325-y}
}

@article{tanaka2025muon_password_gen,
  author  = {H. K. M. Tanaka and others},
  title   = {Autonomous password generation and setting system with cosmic coding and transfer ({COSMOCAT}) and cosmic time calibrator ({CTC})},
  journal = {Sci. Rep.},
  volume  = {15},
  month   = feb,
  year    = {2025},
  note    = {Art. no. 5378},
  doi     = {10.1038/s41598-025-87007-6}
}

@book{menezes2018,
  author    = {A. J. Menezes and P. C. Van Oorschot and S. A. Vanstone},
  title     = {Handbook of Applied Cryptography},
  address   = {Boca Raton, FL, USA},
  publisher = {CRC Press},
  year      = {2018}
}

@techreport{fips197,
  author      = {{National Institute of Standards and Technology}},
  title       = {Advanced Encryption Standard ({AES})},
  institution = {National Institute of Standards and Technology},
  address     = {Gaithersburg, MD, USA},
  number      = {FIPS 197-upd1},
  month       = may,
  year        = {2023},
  doi         = {10.6028/NIST.FIPS.197-upd1}
}

@techreport{rfc5869,
  author      = {H. Krawczyk and P. Eronen},
  title       = {{HMAC}-based extract-and-expand key derivation function ({HKDF})},
  institution = {Internet Engineering Task Force},
  type        = {RFC},
  number      = {5869},
  month       = may,
  year        = {2010},
  doi         = {10.17487/RFC5869}
}

@article{needleman1970,
  author  = {S. B. Needleman and C. D. Wunsch},
  title   = {A general method applicable to the search for similarities in the amino acid sequence of two proteins},
  journal = {J. Mol. Biol.},
  volume  = {48},
  number  = {3},
  pages   = {443--453},
  year    = {1970},
  doi     = {10.1016/0022-2836(70)90057-4}
}

@article{feng2021,
  author  = {J. Feng and others},
  title   = {A thermal bonding method for manufacturing {Micromegas} detectors},
  journal = {Nucl. Instrum. Methods Phys. Res. A},
  volume  = {989},
  year    = {2021},
  note    = {Art. no. 164958},
  doi     = {10.1016/j.nima.2020.164958}
}

@article{wang2025,
  author  = {Y. Wang and others},
  title   = {A high-compression-ratio channel-multiplexing method for micropattern gaseous detectors},
  journal = {IEEE Trans. Nucl. Sci.},
  volume  = {72},
  number  = {3},
  pages   = {765--772},
  month   = mar,
  year    = {2025},
  doi     = {10.1109/TNS.2025.3537159}
}

@article{zhu2019,
  author  = {D. Zhu and others},
  title   = {Development of the front-end electronics for {PandaX-III} prototype {TPC}},
  journal = {IEEE Trans. Nucl. Sci.},
  volume  = {66},
  number  = {7},
  pages   = {1123--1129},
  month   = jul,
  year    = {2019},
  doi     = {10.1109/TNS.2019.2907125}
}

@article{liu2025daq,
  author  = {J. Liu and Y. Wang and C. Feng and S. Liu and Q. Chen},
  title   = {A back-end electronics based on optical fiber communication for small to medium-scale physics experiments},
  journal = {J. Instrum.},
  volume  = {20},
  number  = {5},
  month   = may,
  year    = {2025},
  note    = {Art. no. T05002},
  doi     = {10.1088/1748-0221/20/05/T05002}
}

@article{liu2025sys,
  author  = {S. Liu and others},
  title   = {{FIGURES}: A versatile readout system for high-granularity muography detectors},
  journal = {IEEE Trans. Nucl. Sci.},
  volume  = {72},
  number  = {10},
  pages   = {3433--3441},
  month   = oct,
  year    = {2025},
  doi     = {10.1109/TNS.2025.3595527}
}

@techreport{rukhin2010,
  author      = {A. Rukhin and others},
  title       = {A statistical test suite for random and pseudorandom number generators for cryptographic applications},
  institution = {Nat. Inst. Standards Technol.},
  address     = {Gaithersburg, MD, USA},
  type        = {Special Publication},
  number      = {800-22 Rev. 1a},
  month       = apr,
  year        = {2010},
  doi         = {10.6028/NIST.SP.800-22r1a}
}

@article{tanaka2023ctc,
  author  = {H. K. M. Tanaka},
  title   = {Cosmic time calibrator for wireless sensor network},
  journal = {Sci. Rep.},
  volume  = {13},
  month   = apr,
  year    = {2023},
  note    = {Art. no. 5951},
  doi     = {10.1038/s41598-023-32262-8}
}

\end{document}